\documentclass[a4paper,10pt]{article}
\usepackage{enumerate}
\usepackage{color}
\usepackage[utf8]{inputenc} 
\usepackage[english]{babel}
\usepackage[T1]{fontenc}
\usepackage{graphicx}
\usepackage{amsfonts,amssymb,amsmath,latexsym,amsthm}
\usepackage{textcomp}
\usepackage[pdftex]{hyperref}
\usepackage{geometry}
\title{Full Euler equations for waves generated by vertical seabed displacements}
\author{João Vitor Parada Poletto$^{1}$, David Andrade$^{2}$, Marcelo V. Flamarion$^{3}$ and Roberto Ribeiro-Jr$^{1}$}
\date{}

\begin{document}
\maketitle
\begin{center}
{\footnotesize $^1$UFPR/Federal University of Paran\'a,  Departamento de Matem\'atica, Centro Polit\'ecnico, Jardim das Am\'ericas, Caixa Postal 19081, Curitiba, PR, 81531-980, Brazil  \\
marcelo.flamarion@ufrpe.br }

\vspace{0.3cm}
{\footnotesize 
$^{2}$School of Engineering, Science and Technology, Universidad del Rosario, 111711, Bogot\'{a}, Colombia.

$^{3}$Unidade Acad{\^ e}mica do Cabo de Santo Agostinho, 
	UFRPE/Rural Federal University of Pernambuco, BR 101 Sul, Cabo de Santo Agostinho-PE, Brazil,  54503-900.}



\end{center}


\begin{abstract} 

We present a novel numerical method for simulating the generation and propagation of surface gravity waves by vertical seabed displacements. The cornerstone of our method is the computation of a time dependent conformal map which incorporates the time dependent geometry of the seabed and the wave profile along the free surface. This enables us to handle general geometric configurations of the seabed and the wave. As benchmark we reproduce the results of Hammack on tsunami generation and propagation. Our results show that Hammack's linear theory accurately predicts wave generation. However, as the velocity of the sea bed displacement increases, nonlinear effects become increasingly noticeable. Notably, when the seabed uplift occurs rapidly, the following nonlinear dynamics of the wave differ significantly from the linear dynamics usually associated with tsunami propagation. 
\end{abstract}

\section{Introduction}	

In this article we study the generation and subsequent propagation of surface gravity waves by displacements of the underwater seabed.  We assume an ideal fluid bounded above by a free surface and below by a rigid boundary which follows a prescribed motion. The fluid is initially at rest and surface waves are generated as a result of the geometric deformations of the fluid domain.


The most important applications of this problem is the modelling of seismically generated waves such as tsunamis and their propagation in order to gain insights into tsunami characteristics following seismic events and their propagation.



The literature on mathematical studies of wave generation and propagation due to seabed displacements is vast and gained a lot of attention after the 2004 Indian Ocean earthquake that triggered a tsunami off the coast of Sumatra. One of the earliest references in this topic is the pioneering work of Hammack \cite{Hammack}, who derived a linearized set of equations to model wave generation and successfully compared his theoretical predictions with laboratory experiments. Hammack's work started an extensive investigation into the generation of surface gravity waves using the linearized Euler equations as the basic model. Substantial amount of work on linear generation and propagation of waves was followed by many authors in particular by Dutykh and collaborators, see \cite{Dutykh_2007,Dutykh_2009,Dutykh_2013}. Hence the next step is the investigation of nonlinear effects in the generation and subsequent propagation of surface waves resulting from the seabed motion.

Recent studies on nonlinear waves generated by the deformation of the seabed has primarily been conducted within the framework of asymptotic theory (as discussed in \cite{Michele:2022}),  or by means of Reynolds-Averaged Navier-Stokes equations coupled with a $k-\varepsilon$ turbulence model \cite{Qi_2017,Shen_2022}. Michele et al. \cite{Michele:2022}  observed that weakly nonlinear models predict waves with higher crests and deeper troughs compared to the well-established linear solutions. Furthermore, Qi et al. \cite{Qi_2017} and Shen et al. \cite{Shen_2022}  noticed that when the sea bottom displacement is high enough and rises rapidly, the nonlinear effects in the formation of the wave can not be neglected. Differences between waves generated by linear and nonlinear models demand further investigation into comprehensive nonlinear models. This demand gains additional significance due to the practical applications in engineering and oceanography of the problem at hand.

Our goal is to derive a numerical method which solves the full nonlinear Euler equations for an irrotational flow with two moving boundaries: the free surface and the seabed. The seabed displacement is given by a section of the bed moving vertically either up or down. Our method relies on the computation of a time dependent conformal map which is used to replace the time dependent fluid domain to a uniform strip called the canonical domain. 

The application of conformal maps to simulate fluid dynamics has proven to be successful and has been used in a variety of different problems, including steady free surface over a flat seabed \cite{Dyachenko:2019, Ribeiro:2017},  linear wave propagation over variable topography \cite{FRN}, time dependent free surface waves in channels with infinity depth \cite{Milewski:2010}), and time dependent free surface waves over  spatially variable seabed \cite{Flamarion:2019} and an extension to three dimensional flow by Andrade and Nachbin \cite{AndradeNachbin:2018}. To the best of our knowledge this is the first time that the conformal mapping technique is implemented with two time-dependent boundaries and no fully nonlinear computations have been performed so far for the generation and subsequent propagation of waves generated by vertical seabed displacements within the framework of the Euler equations. 

We compare our numerical simulations of the full Euler equations with the results obtained from Hammack's linearized model. Our simulations indicate that the linear theory effectively predicts the wave generation but falls short in accurately  capturing  the  wave propagation for impulsive displacements. This limitation becomes more prominent as the seabed uplift velocity is increased -- in this case, the dynamics captured by the linear model are entirely different from those of the nonlinear model.

This article is organized as follows: The governing equations, the  conformal mapping technique and its numerical implementation  are presented in Section 2. We present the numerical experiments in Section 3 followed by our final conclusions and considerations.


\section {Formulation}

We consider an ideal fluid whose flow is incompressible and irrotational over a channel of finite depth. Let $h_0 + h(x,t)$ denote the time-dependent seabed and  ${\zeta}(x, t)$ the free surface where $h_0$ is a typical depth. By using $h_0$ as the horizontal and vertical length scale, $(gh_0)^{1/2}$ as a velocity scale, $(h_0/g)^{1/2}$ and as the time scale yields the following form of dimensionless Euler equations

        \begin{align}
         \label{eq:eu1}
            & \Delta{{\phi}}= 0 \;\  \mbox{for} \;\ -1+h(x,t) < y < \zeta (x,t), \\
            \label{eq:eu2}
            & {{\phi}}_{y} =h_{t}+{\phi}_{x}h_{x} \;\ \mbox{at} \;\ y = -1+h(x,t), \\
            \label{eq:eu3}
            & {\zeta}_{t} + \phi_x{{\zeta}}_{x}-{{\phi}}_{y}=0
            \;\ \mbox{at} \;\ y = \zeta (x,t), \\
            \label{eq:eu4}
            & {{\phi}}_{t}+  \frac{1}{2}(\phi_x^2 + \phi_y^2)+{{\zeta}}= 0\;\ \mbox{at} \;\ y = \zeta (x,t).
               \end{align}
We supplement these equations with periodic boundary conditions of period $2L$. 

Equations \eqref{eq:eu1}-\eqref{eq:eu4} are solved numerically by means of the conformal mapping method which is presented next.

\subsection{Conformal mapping for a time-dependent seabed}\label{section:conformal}

        We compute a time-dependent conformal mapping $f$ 
        \begin{equation}\label{eq:direta}
        f(\xi+i\eta,t) = x(\xi,\eta,t)+iy(\xi,\eta,t),
        \end{equation}
        to flatten the free surface and seabed onto a strip of height $D$. Its components $x(\xi,\eta,t)$ and $y(\xi,\eta,t)$ are harmonic functions of $\xi$ and $\eta$ and the imaginary part of the mapping satisfies the boundary conditions 
        \begin{equation} \label{eq:boundaryY}
        y(\xi,0,t)= \mathbf{Y}(\xi,t) \;\ \mbox{and} \;\ y(\xi,-D,t)=-1+\mathbf{H}(\xi,t),
        \end{equation}
        where $\mathbf{H}(\xi,t)=h(x(\xi,-D,t),t)$ contains the information about the bottom corrugations and $\mathbf{Y}(\xi,t)$ contains the information about the instantaneous free surface elevation.

        The function $y$ can be written in Fourier series as 
        \begin{align}\label{eq:y}
        \begin{aligned}
        y(\xi,\eta,t) & = \mathcal{F}^{-1}_{k_j\ne 0}\bigg[\frac{-\coth(k_jD)\sinh(k_j\eta)\widehat{\mathbf{H}}}{\cosh(k_jD)}\bigg] + \frac{1-\widehat{\mathbf{H}}(0,t)}{D}\eta. \\
         & + \mathcal{F}^{-1}_{k_j\ne 0} \bigg[\frac{\sinh(k_j(D + \eta))\widehat{\mathbf{Y}}}{\sinh(k_jD)}\bigg] + \frac{(\eta + D)\widehat{\mathbf{Y}}(0,t)}{D},
         \end{aligned}
         \end{align}
    and from the Cauchy-Riemann equations ($x_\xi=y_\eta$) we can write the function $x$ as 
        \begin{align}\label{eq:x}
        \begin{aligned}
                 x(\xi,\eta,t) & = \mathcal{F}^{-1}_{k_j\ne 0}\bigg[\frac{i\coth(k_jD)\cosh(k_j\eta)\widehat{\mathbf{H}}}{\cosh(k_jD)}\bigg] + \frac{1-\widehat{\mathbf{H}}(0,t)}{D}\xi \\
                 & + \mathcal{F}^{-1}_{k_j\ne 0}\bigg[\frac{-i\cosh(k_j(D+\eta))\widehat{\mathbf{Y}}}{\sinh(k_jD)}\bigg] + \frac{\widehat{\mathbf{Y}}(0,t)}{D}\xi.
                 \end{aligned}
            \end{align}
        
        We are denoting the Fourier coefficients by 
        \begin{align}
            \mathcal{F}_{k_j}[g(\xi)]=\hat{g}(k_j)=\frac{1}{2L}\int_{-L}^{L}g(\xi)e^{-ik_j\xi}\,d\xi,
        \end{align}
        and the inverse Fourier transform by 
        \begin{align}
            \mathcal{F}^{-1}_{k_j}[\hat{g}(k_j)](\xi)=g(\xi)=\sum_{j=-\infty}^{\infty}\hat{g}(k_j)e^{ik_j\xi},
        \end{align}
        where $k_j=(\pi/L)j$, $j\in\mathbb{Z}$.

               The conformal mapping itself is also a periodic function of $\xi$ and we can adjust its horizontal period to match that of the physical problem by choosing an appropriate value of $D$. Indeed we can set
         \begin{equation}\label{eq:xL}
        \int_{-L}^{L} x_\xi(s,\eta,t) \,ds = x(\xi = L,\eta,t) -    x(\xi =- L,\eta,t) = 2L,
        \end{equation}
        which from equation \eqref{eq:x}, reduces to choose the height of the strip as 
        \begin{align}
            D = 1 - \widehat{\mathbf{H}}(0,t) +  \widehat{\mathbf{Y}}(0,t).
        \end{align}
        
        Let $\mathbf{X}(\xi,t)$ be  the horizontal  coordinate of the conformal mapping at $\eta=0$ and $\mathbf{X}_{b}(\xi,t)$ be its trace along the bottom $\eta=-D$. From (\ref{eq:x}) we have
            \begin{align}\label{xxi}
         \mathbf{X}(\xi,t) & = \xi+ \mathcal{F}^{-1}_{k_j\ne 0}\bigg[\frac{i\coth(k_jD)\widehat{\mathbf{H}}}{\cosh(k_jD)}\bigg]  +\mathcal{F}^{-1}_{k_j\ne 0}\bigg[-i\coth(k_jD)\widehat{\mathbf{Y}}\bigg]
          \end{align}
        and

             \begin{align}\label{xbxi}
             \begin{aligned}
        \mathbf{X}_{b}(\xi,t)& =\xi + \mathcal{F}^{-1}_{k_j\ne 0}\bigg[i\tanh(k_jD)\widehat{\mathbf{H}}\bigg] \\
        &+ \mathcal{F}^{-1}_{k_j\ne 0}\bigg[i\coth(k_jD)\bigg[\frac{\widehat{\mathbf{H}}}{\cosh^2(k_jD)} - \frac{\widehat{\mathbf{Y}}}{\cosh{(k_jD)}} \bigg]\bigg].  
        \end{aligned}
          \end{align}
       Note that equation $(\ref{xbxi})$ defines $\mathbf{X}_{b}(\xi,t)$ implicitly since $\mathbf{H}(\xi,t) = h(\mathbf{X}_b,t)$.

Following \cite{Flamarion:2021} we compute $\mathbf{X}_b$ through a fixed point iterative scheme of the form 
 \begin{align}
                \begin{cases}
          \mathbf{X}_{b}^{(p+1)}(\xi,t) = \mathcal{I}\left[ \mathbf{H}^{(p)}(\xi,t) \right],  \quad p = 0,1, \cdots \\[5pt]
                \mathbf{H}^{(p)}(\xi,t) = h( \mathbf{X}_{b}^{(p)}(\xi,t),t ), 
                                        \end{cases}
\end{align}
with initial guess
 \begin{equation*}
       \mathbf{H}^{(0)}(\xi,t_l)   = 
          \begin{cases}
            0& \text{ if }   t_l=0,  \\
               \mathbf{H}(\xi,t_{l-1})    &  \text{ otherwise}, 
                                        \end{cases}
         \end{equation*}
where  $\mathcal{I}$ is the operator
\begin{align}
    \mathcal{I} [\cdot] \equiv \xi
      + \mathcal{F}^{-1}_{k_j\ne 0}\bigg[i\tanh(k_jD) \mathcal{F}[\cdot]\bigg] 
     +  \mathcal{F}^{-1}_{k_j\ne 0}\bigg[i\coth(k_jD)\bigg[\frac{\mathcal{F}[\cdot]}{\cosh^2(k_jD)} - \frac{\widehat{\mathbf{Y}}}{\cosh{(k_jD)}} \bigg]\bigg]
     .
\end{align} 
Further details of the method and its accuracy can be found in \cite{Flamarion:2021}.

\subsection{Euler equations in the canonical coordinates}

We use the conformal map as a new coordinate system.  Let $\bar{\phi}(\xi,\eta,t) ={\phi}(x(\xi,\eta,t),y(\xi,\eta,t),t)$  and  $\bar{\psi}(\xi,\eta,t) = {\psi}(x(\xi,\eta,t),y(\xi,\eta,t),t)$ be the velocity  potential and its harmonic conjugate in the new variables $\xi$ and $\eta$ and denote by $\mathbf{\Phi}(\xi,t)$ and $\mathbf{\Psi}(\xi,t)$ their values along $\eta=0$.

In the new coordinate system the bottom boundary condition, given by equation \eqref{eq:eu2}, becomes $\bar{\phi}_\eta = h_{t}x_{\xi}$ at $\eta =-D$. This allows us to write the velocity potential as

        \begin{equation}\label{eq:Pot1}
              \bar{\phi}(\xi,\eta,t) = \mathcal{F}^{-1}_{k_j\ne 0}\bigg[\frac{\cosh(k_j(\eta+D))\widehat{\mathbf{\Phi}}}{\cosh(k_jD)}
        +\frac{\sinh(k_{j}\eta)}{k_{j}\cosh(k_{j}D)}\widehat{h_{t}x_{\xi}}\bigg] + \widehat{\mathbf{\Phi}}(0)  + {\widehat{h_{t}{x}_{\xi}}(0,t)}\eta.
              \end{equation}
               A similar equation for $\bar{\psi}$ can be obtained by noting that at $\eta = -D$,  $\bar{\psi} = -\int_{\xi_0}^{\xi} h_{t}x_{\xi}d\xi+Q(t)$ where $Q(t)$ is an arbitrary function of $t$ which, without loss of generality, is set to 0. The equation is

        \begin{equation}\label{eq:Pot2}
       \bar{\psi}(\xi,\eta,t) = \mathcal{F}^{-1}_{k_j\ne 0}\bigg[\frac{\sinh(k_j(\eta+D))}{\sinh(k_jD)}\Bigg(\widehat{\mathbf{\Psi}}+\frac{\widehat{\mathbf{I}}}{\cosh(k_{j} D)}\Bigg)
        -\frac{\widehat{\mathbf{I}}\cosh(k_{j}\eta)}{\cosh(k_{j}D)} \bigg] +\frac{\widehat{\mathbf{I}}(0,t)}{D}\eta,\\
              \end{equation}
        where for ease on the notation we write 
         $\mathbf{I}(\xi,t)= \int_{\xi_0}^{\xi} h_{t}x_{\xi}d\xi.$  

         Finally, from the Cauchy-Riemann equations we obtain the following relation between $\widehat{\mathbf{\Phi}_\xi}$ and $\widehat{\mathbf{\Psi}_\xi}$
        \begin{align}\label{PotS}
        \mathbf{\Psi}_{\xi}(\xi,t)=\mathcal{F}^{-1}\Bigg[i\tanh(k_{j}D) \widehat{\mathbf{\Phi}_{\xi}}-\frac{\widehat{h_{t}x_{\xi}}}{\cosh(k_{j}D)} \Bigg],
        \end{align}

        In the new coordinates the kinematic and  dynamic boundary conditions \eqref{eq:eu3}- \eqref{eq:eu4} are

        \begin{align}
                 &\mathbf{X}_{\xi}\mathbf{Y}_t - \mathbf{Y}_{\xi}\mathbf{X}_t = -\mathbf{\Psi}_{\xi}, \label{coupled}\\
                 &\mathbf{\Phi}_t + \mathbf{Y} = - \dfrac{1}{J}\left[-(\mathbf{X}_{\xi}\mathbf{X}_t + \mathbf{Y}_{\xi}\mathbf{Y}_t)\mathbf{\Phi}_{\xi} + (\mathbf{X}_{\xi}\mathbf{Y}_t - \mathbf{Y}_{\xi}\mathbf{X}_t)\mathbf{\Psi}_{\xi} + \dfrac{1}{2}(\mathbf{\Phi}_{\xi}^2 + \mathbf{\Psi}_{\xi}^2)\right], \label{eq:dinamica}
        \end{align}
where $J = \mathbf{X}_{\xi}^2 + \mathbf{Y}_{\xi}^2$ is the Jacobian evaluated at $\eta = 0$. Following \cite{Milewski:2010}  the $\mathbf{X}_t$ and $\mathbf{Y}_t$ dependence in equation \eqref{coupled} can be decoupled by considering the real and imaginary parts of the analytic function $f_t/f_\xi$, evaluated at $\eta = 0$ which yields

        \begin{equation}\label{Sistema}
         Im\left(\frac{f_t}{f_\xi}\right) =  
        \frac{\mathbf{X}_{\xi}\mathbf{Y}_t - \mathbf{Y}_{\xi}\mathbf{X}_t}{J} = -\frac{\mathbf{\Psi}_{\xi}}{J}
        \quad \mbox{and} \quad Re\left(\frac{f_t}{f_\xi}\right) =  
        \frac{\mathbf{X}_{\xi}\mathbf{X}_t + \mathbf{Y}_{\xi}\mathbf{Y}_t}{J} = \mathcal{C}\left[ \frac{\mathbf{\Psi}_{\xi}}{J}\right]
         \end{equation}
                           where $\mathcal{C}\left[ \frac{\mathbf{\Psi}_{\xi}}{J}\right]$  is given by
\begin{equation}
     \mathcal{C}\left[ \frac{\mathbf{\Psi}_{\xi}}{J}\right] = 
             \mathcal{F}^{-1}_{k_j \neq 0}\left[i \coth(k_jD) \widehat{\left[\frac{\mathbf{\Psi}_{\xi}}{J}\right]}\right]  - \widehat{M}(0,t),
\end{equation}
where $M(\xi,t)  = \mathbf{X}_\xi \mathcal{F}^{-1}_{k_j \neq 0}\left[i \coth(k_jD) \widehat{\left[\frac{\mathbf{\Psi}_{\xi}}{J}\right]}\right] +  \mathbf{Y}_\xi \frac{\mathbf{\Psi}_{\xi}}{J} $. Further details about this computation can be found in  \cite{Flamarion:2019}.

        Equations \eqref{Sistema} are solved for $\mathbf{X}_t$ and $\mathbf{Y}_t$ thus obtaining the following equations
        \begin{align}\label{eq_X_Y_t}
            \mathbf{X}_t &= \mathbf{X}_{\xi}\mathcal{C}\left[\dfrac{\mathbf{\Psi}_{\xi}(\xi,t)}{J}\right] + \mathbf{Y}_{\xi}\dfrac{\mathbf{\Psi}_{\xi}(\xi,t)}{J} \\
            \mathbf{Y}_t &= \mathbf{Y}_{\xi} \mathcal{C}\left[\dfrac{\mathbf{\Psi}_{\xi}(\xi,t)}{J}\right] - \mathbf{X}_{\xi}{\left(\dfrac{\mathbf{\Psi}_{\xi}(\xi,t)}{J}\right)}.\label{eq_X_Y_t:Y}
        \end{align}

       Last, by substituting \eqref{eq_X_Y_t} into equation \eqref{eq:dinamica} yields  the following equation for the potential at $\eta = 0$
        \begin{equation}\label{systemYPhi}
                \mathbf{\Phi}_t = \mathcal{C}\left[\dfrac{\mathbf{\Psi}_{\xi}(\xi,t)}{J}\right]\mathbf{\mathbf{\Phi}}_{\xi} - \frac{1}{2J}(\mathbf{\mathbf{\Phi}}_{\xi}^2 - \mathbf{\mathbf{\Psi}}_{\xi}^2) - \mathbf{Y}.
         \end{equation}

Equations \eqref{xxi}, \eqref{PotS}, \eqref{eq_X_Y_t:Y} and \eqref{systemYPhi} are the main result of this section.  They allow us to compute the time dependent conformal map as well as the evolution of the free surface. Those equations are solved numerically from given initial conditions $\mathbf{Y}$ and $\mathbf{\Phi}$ by means of the fourth order Runge-Kutta method and by means of a Fourier spectral discretization for the  variable $\xi$, with all derivatives being  computed  spectrally trough the Fast Fourier Transform (FFT).   Unless mentioned otherwise, we use $N = 2^{12}$ modes in the computation of the FFT and the Runge-Kutta method is used with a time step of $0.01$.

  \section{Numerical experiments}
    We investigate flows generated by vertical displacements of a section of the seabed.  
         For this purpose,  we consider a fluid domain of 200 dimensionless units of length (200 times the depth of the channel).  In the middle of the domain the seabed displacement is given by the    function
                 \begin{equation}\label{eq:h}
          h(x,t) = 
          \begin{cases}
           \dfrac{z_0\left(1 - \cos(\pi t/T)\right) \left(1+\tanh(0.4(b^2 - x^2))\right)}{2\left(1+\tanh(0.4b^2)\right)}
            &
            t \leq T,
            \\[10pt]
            \dfrac{z_0\left(1+\tanh(0.4(b^2 - x^2))\right)}{1+\tanh(0.4b^2)}
            &
            t > T.
          \end{cases}
        \end{equation}

        Note that when $t=T$ the elevation (or depression) of the seabed stops and it reaches its maximum displacement $z_0$. 
        
        Although the seabed disturbance $h(x,t)$ is not a periodic function of $x$, it decays exponentially to zero as $|x| \to \infty$, so we can truncate it to fit the domain $-L\leq x \leq L$, with $L = 100$.        
        We also approximate the boundary conditions by periodic ones.  The main advantage of this setup is that we can directly compare our results with those obtained from the linear model of \cite{Hammack} in the same configuration.


\subsection{Benchmark}\label{sec:benchmark}

 As shown in the seminal work of Hammack \cite{Hammack}, the solution of the linearized version of the  equations (\ref{eq:eu1})-(\ref{eq:eu4}) is 
\begin{equation} 
    \zeta(x,t) =  \mathcal{F}^{-1}\bigg[\text{sech}(k_j)\int_0^t \cos(\omega_j(t-s))\hat{h}_t(k_j,t) ds\bigg],
    \label{eta1}
\end{equation}
where $\omega_j^2 = k_j\tanh{(k_j)}$.
                
The sea bed elevation  \eqref{eq:h} is controlled by three parameters: $z_0$, $b$ and $T$.     
        Following the terminology of \cite{Hammack}, the type of displacement  is impulsive,  when the seabed moves  rapidly ($b/T \gg1$),
        creeping, when the seabed moves slowly  ($b/T \ll1$), and transitional ($b/T  \approx1$). In our simulations an impulsive displacement is achieved with $b = 6.1$ and $T = 0.793$, the transitional displacement with $b = 1.22$ and $T = 1.098$, and the creeping displacement with $b = 0.61$ and $T = 19.154$. These parameters were chosen following the experimental results of Hammack \cite{Hammack}.
        
        Note that the original non-dimensionalization of Hammack takes $\sqrt{gh_0}$, which is the speed of long linear waves, as a reference velocity to define the three regimes. In our case this is taken care of by our choice dimensionless Euler equations.

\begin{figure}
        \centering
               \includegraphics[scale=1.2]{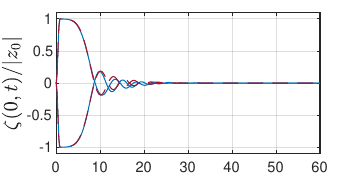}
        \includegraphics[scale=1.2]{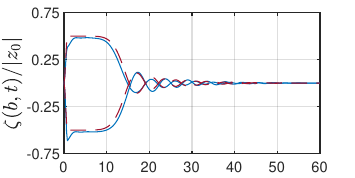}
        \includegraphics[scale=1.2]{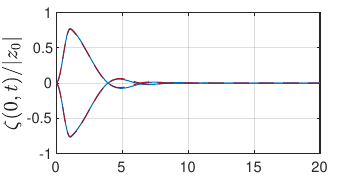}
        \includegraphics[scale=1.2]{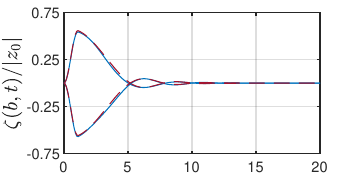}
        \includegraphics[scale=1.2]{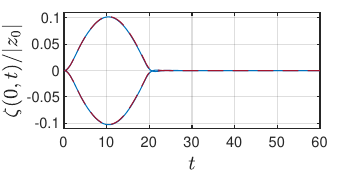}
        \includegraphics[scale=1.2]{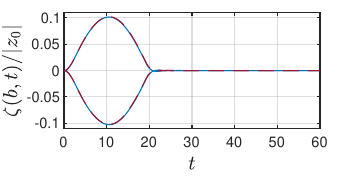}
        \caption{Time series of the wave generation for $|z_0| = 0.1$ in three different regimes: impulsive(top), transitional(middle) and creeping (bottom). Blue solide line: solution of Euler equations. Red dashed line: solution of the linear model of Hammack. }
        \label{Fig:comparasion}
        \end{figure}

        We compare the free surface waves computed from the  system \eqref{eq_X_Y_t:Y}-\eqref{systemYPhi} with those taken from  Hammack's  model, see equation \eqref{eta1}. The resulting simulations are shown in figure \ref{Fig:comparasion} for $|z_0| = 0.1$. The blue solid line shows the solution of the full Euler equations and the red dashed line is the prediction taken from Hammack's model. The plots on the left show the time evolution at $x = 0$, the middle of the disturbance. The plots on the right show the evolution at $x = b$,  around the edge of the seabed deformation. When the seabed moves upwards the  water surface elevates, when it moves downwards we get a depression wave.

\subsection{Similarities and differences: linear vs. nonlinear model}\label{sec:sim_and_dif}

At a glance the time series of the generated wave, see figure \ref{Fig:comparasion}, shows a characteristic symmetric pattern with respect to the $x$ axis; the red lines are indeed always symmetric, a consequence of Hammack's linear approximation. The nonlinear solutions are nearly symmetric in all but the impulsive regime as shown in the upper right panel in figure \ref{Fig:comparasion}. This asymmetry is a nonlinear effect on the generated wave. One can anticipate that such disparities between the linear and nonlinear models will be more pronounced with the increase of $z_0$, i.e. the maximum displacement of the seabed, as will be shown next. 

\begin{figure}
\centering
        \includegraphics[scale=1.2]{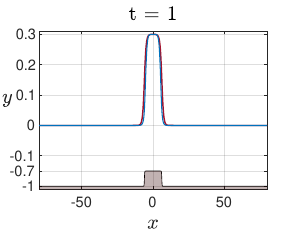}
        \includegraphics[scale=1.2]{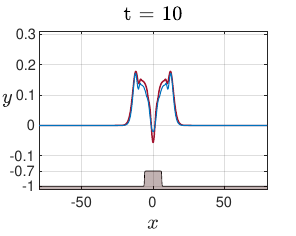}
        \includegraphics[scale=1.2]{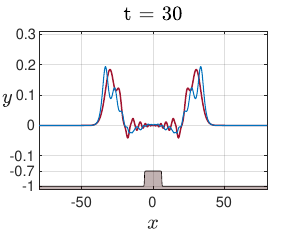}
        \includegraphics[scale=1.2]{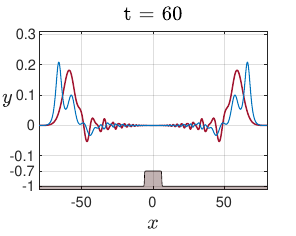}
        \caption{Snapshots of the waves generated for both models for $z_0 = 0.3$. Blue line: solution of the Euler equations. Red line: solution of the linear model of Hammack. Black line: seabed.} 
        \label{Impulsive_z0=0.3}
\end{figure}

We compare the dynamics of the free surface wave when $z_0 = 0.3$, i.e. one third of the depth. Four snapshots of the dynamics are shown in figure \ref{Impulsive_z0=0.3}. For the impulsive regime we use $T = 0.793$ and $b = 6.1$ as before. From the figure one can see that throughout the generation phase, both models behave similarly. Nonlinear effects become visible at a later time; the nonlinear wave propagates faster than the linear one and at $t = 60$ the nonlinear solution is not as smooth as that obtained from Hammack's model. 
A more detailed presentation of the wave dynamics featured in  figure \ref{Impulsive_z0=0.3}.%

\begin{figure}
\centering
        \includegraphics[scale=0.82]{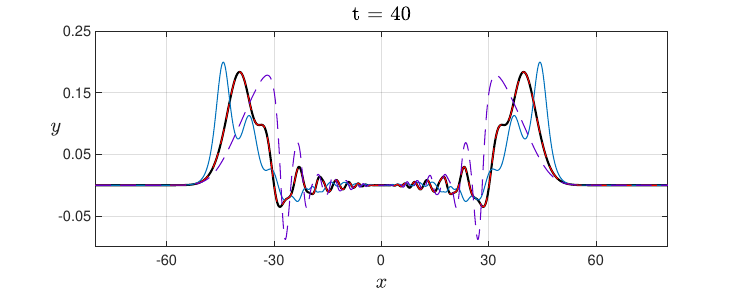}
        \caption{Symmetry comparison with respect to $x$-axis. Blue-solid line refers to the nonlinear solution obtained via the platform lifting and  purple-dashed line represents the opposite of the nonlinear solution via the platform lowering. The linear solutions are represented by the red and black colors.} 
        \label{Comparing symmetry}
\end{figure}

When the seabed submerges the inverse phenomenon happens, the nonlinear solution is smooth and slower than the linear one which is just a reflection along the $x$ axis of the previous one. In order to show the different speeds of propagation we plot the linear and non linear solutions at $t = 40$ for $|z_0| =  0.3 $, see figure \ref{Comparing symmetry}. In the figure the blue solid  line and purple dashed  line are the nonlinear solutions for an upward and downward displacement of the seabed respectively. 
The solution corresponding to the sinking seabed has been reflected around the $x$ axis for comparison. Note that the linear wave (black solid line and red dashed line) lies between the two nonlinear ones. Furthermore one can see that for an uplift motion of the seabed the generated wave is led by a thin higher crest whereas for a sinking seabed the generated wave is trough-led with a smooth but smaller amplitude.

\begin{figure}
\centering
        \includegraphics[scale=1.2]{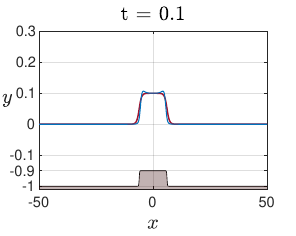}
        \includegraphics[scale=1.2]{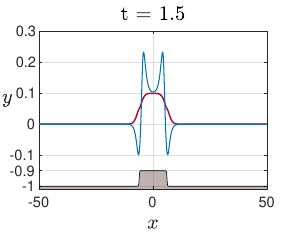}
        \includegraphics[scale=1.2]{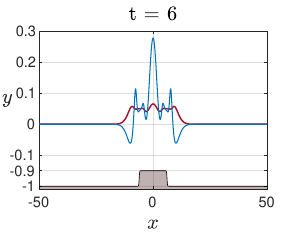}
        \includegraphics[scale=1.2]{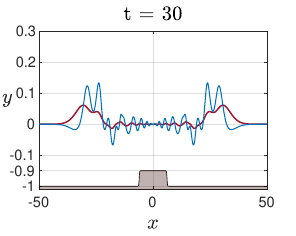}
        \caption{Snapshots of the waves generated for both models in the ``super-impulsive'' movement. Blue line: solution of Euler equations. Red line: solution of the linear model of Hammack. Black line: seabed. Parameters: $\Delta x = 0.02, \Delta t = 2\cdot10^{-4}, N = 2^{13}$ and $z_0 = 0.1$.}
        \label{super_impulsive}
\end{figure}

The last experiment studies wave generation in what we call a ``super-impulsive'' regime, i.e. an extremely fast displacement of the seabed which is achieved by considering $T = 0.01$. 
For the experiments we put $b = 6.1$ and $z_0 = 0.1$ as before. In this regime the seabed moves about eighty times faster than in the impulsive regime.  Snapshots of the generated wave are presented in Figure \ref{super_impulsive}.
Note the difference between the generated linear and nonlinear waves. The latter reaches an amplitude of about three times that of the linear wave. Also, dispersion is enhanced by the nonlinear model as shown by the oscillatory tail above the underwater topography. Last, note that linear wave is crest led whereas the nonlinear one is trough led.
In practical scenarios, the ratio $b/T$ is interpreted as the earthquake rupture speed. For megathrust earthquakes, the rupture speed typically reaches approximately 1.0 km per second \cite{Weng:2022}. Althoug the parameter set used in this experiment may result in unrealistic configurations, it can serve as a means to explore the limits of the full Euler equation.

\section{Conclusion}
\label{Conclusion}

In this work, we introduced a novel numerical method for solving the full Euler equations in the presence of variable spatial and temporal topography, using the conformal mapping technique and spectral numerical methods. To validate our numerical approach, we compared its predictions with solutions obtained from the linear model of Hammack. Generally, the results from both the nonlinear theory and the linear model of Hammack exhibited good qualitative agreement across many regimes of interest. However, the linear model fails to capture the behavior in the super-impulsive regime, where nonlinear terms dominate the wave dynamics. It is noteworthy that the numerical methods presented in this study have versatile applications beyond the specific problem addressed. These techniques can be extended to tackle a wide range of problems, including landslides, interactions of solitary waves with variable-speed obstacles, and other scenarios involving complex topographies. Furthermore, the conformal mapping technique combined with spectral methods opens up opportunities for investigating various fluid dynamics and wave propagation problems in challenging environments, such as coastal regions, underwater structures, and atmospheric conditions. The ability to handle nonlinear dynamics and complex topographies enhances the applicability and significance of the numerical methods presented in the article across multiple scientific disciplines, contributing to a deeper understanding of natural phenomena and aiding in practical engineering and environmental studies.


\section*{Acknowledgments}
M.V.F. and R.R.J. are grateful to IMPA for hosting them as visitors during the
2023 Post-Doctoral Summer Program. R.R.Jr.
The work of    J.V.P.P was support by the  the Coordenação de
Aperfeiçoamento de Pessoal de Nível Superior - CAPES.


\begin{thebibliography}{999}


\bibitem{AndradeNachbin:2018}
{\sc Andrade, D. \& Nachbin, N} 2018 Two-dimensional surface wave propagation over arbitrary ridge-like topographies. SIAM Journal on Applied Mathematics, 78(5), 2465-2490.

\bibitem{Dutykh_2007}
{\sc Dutykh, D \&  Dias, F.} 2007 Water waves generated by a moving bottom. In: Tsunami and Nonlinear waves. Berlin, Heidelberg: Springer Berlin Heidelberg, p. 65-95.

\bibitem{Dutykh_2009}
{\sc Dutykh, D \&  Dias, F.} 2009 Tsunami generation by dynamic displacement of sea bed due to dip-slip faulting. {\it Math. Comput.  Simulat.}, v. 80, n. 4, p. 837-848.


\bibitem{Dutykh_2013}
{\sc Dutykh, D., Mitsotakis, D., Gardeil, X.,  Dias, F.} 2013 On the use of the finite fault solution for tsunami generation problems. {\it Theor. Comp. Fluid Dyn.}, v. 27, p. 177-199.


\bibitem{Dyachenko:2019}
{\sc Dyachenko, S. A., \& Hur, V. M.} 2019 Stokes waves with constant vorticity: folds, gaps and fluid bubbles. {\it J. Fluid Mech.}, 878, 502-521.


\bibitem{Flamarion:2019}
{\sc Flamarion, M. V., Milewski, P. A., \& Nachbin, A.} 2019 Rotational waves generated by current‐topography interaction. {\it Stud. Appl. Math.}, 142(4), 433-464.

\bibitem{FRN}
{\sc Flamarion, M. V., Nachbin, A. \& Ribeiro-Jr, R.} 2020 {Time-dependent {K}elvin cat-eye structure due to current-topography interaction}, {\it J. Fluid Mech.}, {\bf 889}, pp. A11.

\bibitem{Flamarion:2021}
{\sc Flamarion, M. V., \& Ribeiro‐Jr, R.} 2021 An iterative method to compute conformal mappings and their inverses in the context of water waves over topographies. {\it  Int. J. Numer. Meth. Fl.}, 93(11), 3304-3311.



\bibitem{Hammack}
{\sc Hammack, Joseph L.} 1973 {A note on tsunamis: their generation and propagation in an ocean of uniform depth}, {\it J. Fluid Mech.}, {\bf 60}, pp. 769-799.



\bibitem{Michele:2022}
{\sc Michele, S., Renzi, E., Borthwick, A.G.L., Whittaker, C. and Raby, A.C.}, 2022 Weakly nonlinear theory for dispersive waves generated by moving seabed deformation. {\it J. Fluid Mech.}, 937, p.A8.


\bibitem{Milewski:2010}
  {\sc Milewski, P., Vanden-Broeck, J., \& WANG, Z.}  2010 Dynamics of steep two-dimensional gravity–capillary solitary waves. {\it J. Fluid Mech.} , 664, 466-477. 




\bibitem{Qi_2017}
{\sc Qi, M.,  Kuai, Y.,  Li, J.}  2017 Numerical simulation of water waves generated by seabed movement. {\it Appl. Ocean Res.} , v. 65, p. 302-314.


\bibitem{Ribeiro:2017}
{\sc Ribeiro-Jr, R., Milewski, P. A., \& Nachbin, A.} 2017 Flow structure beneath rotational water waves with stagnation points. {\it J. Fluid Mech.} , 812, 792-814.



\bibitem{Shen_2022}
{\sc Shen, Y., Whittaker, C.N., Lane, E.M., Power, W. and Melville, B.W.}  2022 Interference effect on tsunami generation by segmented seafloor deformations. {\it Ocean Eng.}, 245, p.110244.



\bibitem{Weng:2022}
{\sc Weng, H. \&  Ampuero, J.P.}, 2022 Integrated rupture mechanics for slow slip events and earthquakes. {\it Nat.  Commun.}, 13 (1), p.7327.


\end{thebibliography}


\end{document}